\documentclass{kluwer}    

\newdisplay{guess}{Conjecture}

\begin{document}                                                                                   
\begin{article}
\begin{opening}         
\title{On the self-consistency of evolutionary synthesis models}
\author{Miguel \surname{Cervi\~no}}  
\runningauthor{Miguel Cervi\~no}
\runningtitle{Self-consistency of synthesis models}
\institute{IAA (CSIC) Camino bajo de Hu\'etor 24, Granada 18008, Spain\\
LAEFF (INTA) Apdo. 50727, Madrid 28080, Spain}
\date{September 30, 2002}

\begin{abstract}
Evolutionary synthesis models have been used to study the physical
properties of unresolved populations in a wide range of scenarios.
Unfortunately, their self-consistency are difficult to test and there are
some theoretical open questions without an answer: (1) The change of the
homology relations assumed in the computation of isochrones due to the
effect of stellar winds (or rotation) and the discontinuities in the
stellar evolution are not considered. (2) There is no a consensus about how
the isochrones must be integrated. (3) The discreteness of the stellar
populations (that produce an intrinsic statistical dispersion) usually are
not taken into account, and model results are interpreted in a deterministic
way instead a statistical one... The objective of this contribution is to
present some inconsistencies in the computation and some cautions in the
application of the results of such codes.
\end{abstract}
\keywords{Stars: evolution -- Galaxies: stellar content -- open
clusters and associations: general}

\end{opening}           

\section{Introduction}  

In the last few years, the increasingly detailed observations of stellar
populations in a wide variety of environments have provided a huge amount of
high quality data, which have been used to constrain both 
stellar evolution theory and stellar models among many other variables.

In contrast, the development of increasingly complicated evolutionary
synthesis codes has, in general, focused only on the use of updated
physical ingredients, but some of their underlying hypotheses (track
interpolation and isochrone computation in particular) have remained
unchanged. Unfortunately, their self-consistency are difficult to test and
there are some theoretical and practical open questions without an answer.

\section{From tracks to isochrones: Equivalent evolutionary points}

An homology transformation gives us a scale relation of a property $x$ of
stars (L, T$_\mathrm{eff}$, $\tau_\mathrm{ms,life}$) with their Mass, $M$,
mean molecular weight, $\mu$, nuclear energy production, $\epsilon_0$,
opacity coefficient, $\kappa_0$, and Hydrogen abundance in the core, $X$, for
the same configuration. As an example, for the CNO burning, Kramers opacity
law and radiative transport configuration the relations are:

 \begin{eqnarray}
R(M) &\propto &(\epsilon_0  \kappa_0)^{1/20} \mu^{13/20} M^{4/5} \nonumber \\
L(M) &\propto &\kappa_0^{-1} \mu^{4} M^{3} \nonumber \\
\tau_{ms}(M) &\propto &\kappa_0 \mu^{-4} X M^{2} \nonumber 
\end{eqnarray}

Other configurations means, mainly, a change in the exponents. These
relations, together with $L \propto R^2 T_{\mathrm{eff}}^4$, allow us to
define equivalent evolutionary points: points where $\kappa_0$, $\mu$
(structure) and $\epsilon_0$, $X$ (burning state) are similar. Then,
interpolations must be done in the $log M - log x$ plane. Note that the $M$
value used in the homology relations {\it does not refer to the initial
mass}, but the current mass of the star in the equivalent evolutionary
point.

This simple scheme would be a good approximation if (a) the stellar tracks
are close enough, (b) there are no discontinuities in the stellar evolution
(i.e. adjacent tracks ``resemble'' each other), and (c) stellar winds do
not vary the stellar evolution.  Unfortunately, these conditions are not
always true. There are intrinsic discontinuities in the stellar evolution
at low mass (c.f. \citeauthor{TG76} \citeyear{TG76}) and high mass
(Wolf-Rayet, WR, vs. non-WR tracks).  The problem is specially dramatic in
high-mass stars, where the effects of stellar winds strongly affects the
stellar evolution and burning lifetimes of the stars, and may produce
failures in the use of homology relations when the initial mass is
used. Even more, there are no homology relations to compute mass lost rates
(which are, indeed, an input of the evolutionary tracks) neither the
surface abundances (which are related with the internal structure and the
mass lost rate and which are fundamental for obtain the WR population). It
means that, at least in the case of massive stars, where the mass lost
rates affects the stellar evolution, {\it isochrones do not necessary
reflect the physics of the assumed tracks} if they are obtained from the
interpolation of the initial mass (as it is usual in synthesis codes for
starburst galaxies).

\section{From isochrones to integrated properties}

Assuming that the isochrone reflect the input physics used in the
evolutionary tracks, the next step is to populate the isochrone and obtain
the integrated properties. It can be done by a direct convolution of the
isochrone with the Initial Mass Function (IMF)\footnote{Note that the IMF
gives us a {\it probability} of obtain certain amount of stars in a
given mass interval.} and the Star Fromation Rate, together with the
individual stellar properties.

This integration can be done using Monte Carlo simulations (where the
number of stars used in the simulation plays a fundamental role), analytical
integrations (but it is needed an analytical formulation of the isochrone
and the stellar properties), or numerical integrations (where the mass
intervals must be chosen in such a way that all the relevant evolutionary
phases are correctly included in the computation).

Other possible solution is the use of the so-called {\it Fuel Consumption
Theorem, FCT,} \cite{B89}. In this case, the relevant Post-Main Sequence
stages are approximated by a single stellar track of the star with mass at
the turn-off point of the Main Sequence, $M_{TO}$. Note that it is
needed to assure that the isochrone and the $M_{TO}$ stellar track are
similar enough. I refer to \citeauthor{GB98} (\citeyear{GB98}, sect. 2)
or \citeauthor{MG01} (\citeyear{MG01}) for more information.

However, the results of synthesis codes differs qualitatively, depending on
the way how this integration is performed...

\section{The use of synthesis models}

The final question is the comparison of the results of synthesis models
with real data. In this case, it must be taken into account that (1)
synthesis models give us a {\it mean} value of the observed properties
(that is only exact in the asymptotic limit of an infinite number of stars
in the cluster), and (2) the total luminosity of the modeled cluster must
be larger than the individual luminosity of any of the stars included in
the model. This last constrain imposes a minimum amount of stars (or,
equivalently, a minimum initial stellar cluster mass for single stellar
populations or star forming rate for composed ones) where the results of
synthesis models can be compared safely with real data. This limits is
around 10$^5$ M$_\odot$ for a single stellar population using a Salpeter
IMF slope in the range 0.09--120 M$_\odot$ (c.f. Cervi\~no 2002
submitted). Below this limit, the presence of $\pm 1$ star may influence
the resulting properties and sampling effects must be included in the
model. Even more, the resulting properties may show bimodal distributions
and the results of synthesis models may be biased respect the the observed
properties (Cervi\~no \& Valls-Gabaud 2002).  The relevance of these
sampling effects have been illustrated in \citeauthor{Bru01}
(\citeyear{Bru01}) and \citeauthor{CLC00} (\citeyear{CLC00}, see also
\citeauthor{CVGLMH02} \citeyear{CVGLMH02} and references therein for an
analytical formalism).  These sampling effects may also play an important
role in chemical evolution models, as it is shown in \citeauthor{CM02}
(\citeyear{CM02}).

\section{Conclusions}

It is needed a more careful (physical) study in the way how interpolations
between tracks are done, specially for massive stars (WR).  There are also
some open questions that, at this moment, had not been solved
theoretically, as example: What is the scheme needed to obtain realistic
results for systems with metallicities that differs from the one tabulated
in tracks and isochrones? How homology relations works in rotating stars?

From the observational point of view, any comparison of the results of
synthesis models with real data must take into account that the synthesis
model results are a BAND, with a intrinsic dispersion that depends on the
cluster stellar mass, instead an infinitely narrow line.  Evermore, there
is an intrinsic limit in the cluster stellar mass below any comparison of
observed data with synthesis models must be performed including sampling
effects in the model.

\acknowledgements Useful comments have been provided by Valentina Luridiana
and Enrique P\'erez. This project has been partially supported by the AYA
3939-C03-01 program. I also thanks the CE for financial support for
attending the conference.


\theendnotes

\end{article}
\end{document}